\documentclass[letterpaper,twocolumn,10pt]{article}
\usepackage{usenix}
\usepackage{graphicx}
\usepackage{booktabs}
\usepackage{array}
\usepackage{xspace}
\usepackage{amsmath}

\newcommand{\mlmodel}{\texttt{.mlmodel}\xspace}
\newcommand{\mlmodelc}{\texttt{.mlmodelc}\xspace}
\newcommand{\mlpackage}{\texttt{.mlpackage}\xspace}
\newcommand{\mlprogram}{\texttt{mlProgram}\xspace}
\newcommand{\coremltools}{\texttt{coremltools}\xspace}
\newcommand{\zwsp}{\mbox{ZWSP}\xspace}
\newcommand{\zwj}{\mbox{ZWJ}\xspace}
\newcommand{\zwnj}{\mbox{ZWNJ}\xspace}

\begin{document}

\title{Harvesting Apple's Orchard: A Security Analysis of On-Device CoreML Models Across the Apple Ecosystem}

\author{
{\rm Justice Owusu Agyemang}, {\rm Winifred Harriet Asante}, \\
{\rm Kwame Opuni-Boachie Obour Agyekum}, {\rm Kwame Agyeman-Prempeh Agyekum}, \\
{\rm Francisca Adoma Acheampong}, {\rm Jerry John Kponyo} \\
VIA Cybersecurity Lab, Kwame Nkrumah University of Science and Technology, Kumasi, Ghana \\
}

\maketitle

\begin{abstract}
Apple ships hundreds of task-specific machine-learning models inside iOS,
iPadOS, macOS, and tvOS, mediating spam filtering, battery prediction,
suggestions, and media analysis.
We present, to our knowledge, the first systematic security analysis of
this on-device model surface.
We harvested 1{,}137 CoreML artifacts from six platform images (iOS~26.0,
iPadOS~26.0.1, an iOS~26.1 firmware add-on set,
tvOS~26.6, iOS~12.5.8, and macOS~26.5) and two device harvests (an
iPhone~X on iOS~16.7 and a jailbroken iOS~26.1 virtual test device); on
the test device alone we inventoried 320 model files across 54 framework
groups, and a conversion pipeline recovered 155 previously opaque bundles
as loadable \mlpackage{} archives. Eight findings.
(1)~368 of 1{,}137 artifacts expose learned parameters in plaintext (281
blobs, 369.61~MiB) with no encryption in any container our analyzer
recognizes, and 166 of 166 MIL bundles leak internal model-database tokens.
(2)~CoreML specification versions 1--9 coexist in shipped artifacts,
and locale variants diverge architecturally.
(3)~Across 82{,}092 adversarial evaluation records (221 model paths;
77{,}396 \texttt{predict()} calls with a comparable output), worst-case
flip rates reach 100\%; zero-width spaces flip 65\% (13 of 20)
of the Indian-English SMS filter and 70\% (14 of 20) of the US-English
variant under a surrogate embedding for its unresolved production
tokenizer.
(4)~On the test device, a 16-byte weight edit flips 7 of 8 probe inputs of
the interaction-prediction model (the eighth a 50/50 tie); no runtime
integrity check was observed, and on a jailbroken device one mount
suspends the seal.
(5)~Loading Apple's own hand-gesture model crashes \texttt{libBNNS} through
the public CoreML API (4/4 on production M2 hardware; also reproduced on
macOS~27.0)---a robustness defect from a 378-row weight walk with no
observed bound check. Apple verified the behavior and classified it as a
self-crash that does not cross a security boundary; we agree, and treat it
as an availability/robustness finding rather than a vulnerability.
(6)~Eleven shipped bundles carry \texttt{isUpdatable=1} update hooks.
(7)~On the virtualized test device, SMS text-classifier latency is
13.1--52.0~ms (2.3--3.6~ms tokenized, under 0.2~ms legacy-format), with
92.4\% of text-classifier time in the CPU-side NLP core (tokenizer,
embedding, and network), not ANE execution.
(8)~The SMS classifiers label 8--9 of 10 hand-crafted phishing messages
(80--90\%) as Transactional---the model's own output, before any
manipulation; we measure classifier labels, not inbox-placement outcomes.
We filed three reports with Apple Product Security Security: the crash, the internal
build-token leakage, and the internal storage-path disclosure. Apple
reproduced the crash but declined all three as findings that do not cross a
security boundary; we report them, and the weight-integrity and phishing
findings (which we did not file), as robustness results.
\end{abstract}

\section{Introduction}
\label{sec:intro}

Apple devices ship hundreds of on-device models that run silently in the
background: they decide whether an SMS is promotional or transactional,
predict when to finish charging, rank contacts, apps, and widgets, and
analyze every photo and video in the library.
They are a security boundary in the same sense as a system daemon: they
consume untrusted input (messages, media, feature vectors from other
processes), other components treat their outputs as ground truth, and the
platform vendor ships and updates them, not the user.

Unlike Apple's cloud services, this surface has received almost no
systematic public study: the models ship as compiled \mlmodelc{} bundles
whose serialization is not documented and are scattered across dozens of
private framework directories under \texttt{/System/Library}; extracting
them programmatically requires firmware images or a jailbroken device.
Prior work studies application-embedded models~\cite{mobile_audit} and
API-based extraction~\cite{model_extraction}, not the models the OS itself
executes with system privileges.

This paper closes that gap with a full-system census and security analysis
of Apple's CoreML model surface.
We harvest firmware and restore images (IPSW, Apple's signed
device-firmware archives) with the \texttt{ipsw} tool~\cite{ipsw_tool}, an
iPhone~X running iOS~16.7.10/16.7.11, and a jailbroken iOS~26.1 virtual test
device (a local macOS~26.5 \texttt{/System/Library} harvest is inventoried
outside the corpus).
We (i)~run a static analyzer over every artifact, (ii)~recompile native
\mlmodelc{} bundles into loadable \mlpackage{} archives, and (iii)~execute
four campaigns: accuracy and adversarial robustness, latency, weight
poisoning, and crash testing.

\paragraph{Contributions.}
\begin{enumerate}
    \item \textbf{Census.}
    We inventory 320 model files across 54 framework groups on the test
    device and catalog 1{,}137 artifacts (98 \mlmodel{} protobufs,
    909 compiled \mlmodelc{} bundles, 130 loose blobs) from eight source
    groups---six platform images and two device harvests.
    All 22 iPadOS protobufs are SHA-256-identical to their iOS counterparts,
    and 86 of tvOS's 102 artifacts---84 of its 100 bundles (51 byte-identical
    plus 33 metadata-identical), plus both protobufs---have an identical
    counterpart in the other firmware harvests (Section~\ref{sec:census}).
    \item \textbf{Unlocking compiled bundles.}
    A three-part toolchain recompiles native \mlmodelc{} bundles into
    loadable \mlpackage{} archives, recovering 155 models---73 of 276 bundles
    on the test device and 82 of 579 from firmware
    (Section~\ref{sec:conversion}).
    \item \textbf{Format and architecture analysis.}
    We characterize the specification-version distribution (v1--v9 in
    production), the four weight-container formats, and the architectural
    divergence between locale variants of the SMS filter, including an
    8-bit-quantized Siri wake-word transformer recovered from the iOS~26.1
    image extract (Section~\ref{sec:formats}).
    \item \textbf{Adversarial evaluation.}
    82{,}092 adversarial evaluation records across 221 model paths
    (77{,}396 \texttt{predict()} calls with a comparable output,
    Section~\ref{sec:adversarial})
    measure worst-case flip rates up to 100\%, with zero-width character
    insertion flipping 65\% (13 of 20) of the best-performing SMS filter
    and feature-space attacks flipping up to 67\% on tree-ensemble models
    (Section~\ref{sec:adversarial}).
    \item \textbf{Weight exposure.}
    368 of 1{,}137 artifacts expose learned parameters in plaintext
    (281 blobs, 369.61~MiB) with no encryption in any container our analyzer
    recognizes and no runtime integrity check observed, and 166 of 166 MIL
    bundles leak Apple's internal build tokens (Section~\ref{sec:weights}).
    \item \textbf{On-device poisoning and the updatable surface.}
    On the test device, a 16-byte edit to one weight file flips 7 of 8
    probe inputs (the eighth becomes a 50/50 tie) of the system's
    interaction-prediction model; ten further bundles ship
    \texttt{isUpdatable=1} hooks that treat runtime weight replacement as an
    intended interface, unseparated from tampering
    (Sections~\ref{sec:poisoning} and~\ref{sec:updatables}).
    \item \textbf{A deterministic crash in a shipped model.}
    Loading Apple's own \texttt{hand\_gesture} model through the public
    CoreML API segfaults inside \texttt{libBNNS} (4/4 on production M2
    hardware; also reproduced on macOS~27.0); we trace the fault to a 378-row weight walk
    that performs no observed bound check; we assess it as a
    robustness defect, not an exploitable vulnerability
    (Section~\ref{sec:crash}).
    \item \textbf{Measured performance envelope.}
    Device-level numbers replace workstation proxies: 13.1--52.0~ms for SMS
    text classifiers, 2.3--3.6~ms for raw network variants, $<$0.2~ms for
    the legacy-format model, with $\sim$92.4\% of classifier latency in the
    NLP core, not the ANE (Section~\ref{sec:latency}).
\end{enumerate}

\paragraph{Ethics and disclosure.}
All models were harvested from devices and firmware images we own, using a
public jailbreak on a research virtual device and publicly distributed
firmware.
No user data was collected or inspected.
Model weights are Apple's intellectual property: we publish metadata,
analyses, and attack descriptions, and we do not redistribute weights.
The deterministic crash, the internal build-token leakage, and the internal
storage-path disclosure were filed with Apple Product Security Security (tracking ID
OE1104416693631 for the crash); Apple reproduced the crash but declined all
three as findings that do not cross a security boundary
(Section~\ref{sec:disclosure}).

\section{Background}
\label{sec:background}

\subsection{CoreML model formats}

CoreML~\cite{apple_coreml} is Apple's on-device inference framework (iOS~11
onward). Three formats appear in this study:
\mlmodel{} is a self-contained Protocol Buffers file holding the
architecture, weights, input/output schema, and metadata; it is the
interchange and archival format, loadable with
\coremltools~\cite{coremltools} and produced by Create~ML~\cite{apple_create_ml}.
\mlmodelc{} is the compiled runtime bundle the system executes: a directory
of \texttt{coremldata.bin}, \texttt{metadata.json}, weight files, and
optional analytics, with no public decompiler; Xcode builds it from
\mlmodel{} sources~\cite{apple_compiling}.
\mlpackage{} separates architecture and weights into directory components
and is the modern distribution format~\cite{wwdc2021_tune}; our conversion
pipeline (Section~\ref{sec:conversion}) emits it.

Since iOS~15, Apple has migrated models to the \mlprogram{} model type,
encoding the network as a program in Apple's Model Intermediate Language
(MIL)~\cite{wwdc2021_tune} (first shipped for converters in
2020~\cite{wwdc2020_coreml}); MIL bundles lower at runtime through Apple's
Espresso compiler to E5RT (CPU kernels backed by BNNS) or the Neural
Engine (ANE).

\subsection{Distribution and update paths}

Models reach a device by three routes: \emph{OS-embedded} models ship in
the signed system volume and update only with the OS; \emph{TrialAssets}
are an A/B asset mechanism packaging variants independently of the OS
image; \emph{MobileAsset (AssetsV2)} bundles download on demand outside the
system volume (\texttt{/System/Library/AssetsV2/} on macOS,
\texttt{/var/MobileAsset/} on iOS). CoreML also supports on-device
personalization: a bundle compiled with \texttt{isUpdatable} exposes an
\texttt{update}\allowbreak\texttt{Parameters/} directory and can be retrained at runtime
through \texttt{MLUpdateTask}~\cite{mlupdatetask}.

\subsection{The sealed system volume}

Modern Apple platforms boot from a signed, sealed APFS snapshot: the root
volume is read-only at runtime and its seal is verified at boot, so
modifications under \texttt{/System} do not survive a
reboot~\cite{apple_platform_security}. Platform code does not re-verify
individual data files at use time; Section~\ref{sec:poisoning} measures
what that implies for models.

\subsection{Threat model}
\label{sec:threat}

We consider three adversary classes, ordered by capability.

\textbf{A1 --- sandboxed application (no root).}
An app can call the public CoreML API on any model it ships, request that
the runtime load a model from a path it can read (including system model
paths the sandbox exposes), and feed inputs to system models through public
framework APIs (e.g., photo-library additions, feature vectors); it cannot
write the Messages database or modify \texttt{/System}.

\textbf{A2 --- local root.}
\emph{A2a (runtime root on a stock device):} the adversary reads every
model file and writes non-sealed paths, but cannot make the sealed system
volume writable.
\emph{A2b (jailbreak / boot-chain control):} the adversary additionally
suspends the volume seal for the current boot
(Section~\ref{sec:poisoning}) and persists modifications across reboots;
demonstrations that require writing under \texttt{/System} are A2b unless
stated otherwise.

\textbf{A3 --- remote adversary.}
A remote adversary supplies content---SMS text, MMS or email payloads,
media---that a system model consumes; the weakest capability, but the one
that matters most for user-visible features such as spam filtering and
media analysis.

Out of scope: malicious models or weights injected through Apple's own
delivery channels (firmware and signed assets are assumed authentic),
converter-toolchain supply-chain compromise, network interception of
MobileAsset downloads, and malicious Apple insiders.

We ask four questions of these classes:
\textbf{RQ1}~what models ship and what they
contain (Sections~\ref{sec:census}--\ref{sec:formats});
\textbf{RQ2}~their behavior under adversarial input
(Section~\ref{sec:adversarial});
\textbf{RQ3}~what reading or writing model bytes achieves
(Sections~\ref{sec:weights}--\ref{sec:updatables});
\textbf{RQ4}~the on-device performance envelope
(Section~\ref{sec:latency}).

\section{Methodology}
\label{sec:method}

\subsection{Harvesting}
\label{sec:harvesting}

We harvest models from three source families:

\paragraph{Firmware and restore images.}
We download images for iOS~26.0 (iPhone~17 Pro, 23A341),
iPadOS~26.0.1 (iPad Pro M5, 23A8464), tvOS~26.6, iOS~12.5.8, and
macOS~26.5, and mount them with \texttt{ipsw mount fs}~\cite{ipsw_tool},
which copies files without the byte-pattern rewriting that breaks protobuf
sources during archive extraction (every \mlmodel{} file recovered this way
loads cleanly).
A \texttt{find} pass over the mounted volumes collects every
\texttt{*mlmodel*} path.
A separate extraction pass over the iOS~26.1 image (August 2026) adds 19
artifacts beyond the earlier harvests: 14 voice-trigger instances (9
\texttt{hsjs}-variant and 5 plain), the three wireless-resource-management
models, and the two USEN SMS-filter protobufs. (All three families also
ship elsewhere: the voice-trigger models appear in the iOS~26.0,
iPadOS~26.0.1, and tvOS~26.6 harvests; the WRM models in the iOS~26.0 and
iPadOS~26.0.1 harvests and on the iOS~26.1 test device; and the USEN
protobufs in the iOS~26.0 and iPadOS~26.0.1 harvests, the iPhone~X
harvest, and the iOS~26.1 test device.)

\paragraph{Device harvest.}
A direct harvest from an unlocked iPhone~X running iOS~16.7.10/16.7.11
(July 2026) contributes 30 artifacts---21 Duet suggestion models,
6 SmsFilter protobufs (four of which are byte-identical to files that
still ship on the iOS~26.1 test device), and 3 private SMS models---in
the legacy formats that precede 26.x.

\paragraph{Test device.}
We run iOS~26.1 (23B85) on a research virtual device---a
vphone-cli VM presenting an iPhone99,11 profile with 8~CPU cores, 8~GB of
memory, and NAT-only networking---under palera1n-rootless~\cite{palera1n},
which provides a root shell while the system volume remains sealed.
The virtual device is the only source with OS versions newer than the
public IPSW set and with MobileAsset, TrialAssets, and post-setup assets
present; throughout, ``the test device'' means this virtualized instance,
not production hardware.
Enumeration again uses \texttt{find} over
\texttt{/System/Library}, \texttt{/var}, and the asset stores.

\paragraph{macOS.}
The macOS~26.5 image contributes the 17 \mlmodel{} protobufs counted in
the corpus.
A separate harvest of a local (non-image) macOS~26.5
\texttt{/System/Library} finds 361 \mlmodelc{} directories (209 with
parseable \texttt{metadata.json}) and 102 AssetsV2 bundles;
Figure~\ref{fig:macos}b shows the model-type distribution of the 209
parseable bundles.
This harvest stays outside the 1{,}137-artifact record set, which counts
only the 17 macOS protobufs.

\subsection{Unlocking compiled bundles}
\label{sec:conversion}

Native \mlmodelc{} bundles are not loadable by \coremltools{} because they
carry no \texttt{Manifest.json}---a file the desktop toolchain expects but
the on-device runtime never needs. We built a three-stage pipeline:
(i)~\texttt{mlmodelc\_loader} parses \texttt{coremldata.bin} and recovers
the MIL program text and weight blobs;
(ii)~\texttt{mil\_text\_rewrite} normalizes the recovered MIL---renaming
opset symbols to CoreML\texttt{\{spec-1\}} names, coercing integer shapes to
\texttt{int64}, and reconciling the model description---and emits a
\mlpackage{} directory;
(iii)~\texttt{verify\_bundles} loads every emitted package in a subprocess
and runs a prediction, so that a compiler crash cannot take down the
campaign.

\begin{table}[t]
\centering
\caption{Conversion outcomes for the two native bundle corpora.
``Converted'' bundles load and predict; ``native'' bundles carry no MIL and
use the native runtime path; ``predict'' bundles compile but fail at
prediction; ``crashed'' bundles signal the compiler.
All four prediction failures are \texttt{vmtracker}; the seven compiler
crashes are \texttt{MentionGenerationModel} and \texttt{hand\_gesture}.}
\label{tab:conversion}
\scriptsize
\setlength{\tabcolsep}{4pt}
\begin{tabular}{@{}lrr@{}}
\toprule
\textbf{Outcome} & \textbf{Live iOS 26.1} & \textbf{Firmware} \\
\midrule
Bundles examined      & 276 & 579 \\
Converted (load+predict) & \textbf{73} & \textbf{82} \\
Native OK (no MIL)    & 163 & 403 \\
Native fail (no MIL)  & 37  & 86 \\
Predict failure       & 1 (\texttt{vmtracker}) & 3 (\texttt{vmtracker}) \\
Compiler crash        & 2 & 5 \\
\bottomrule
\end{tabular}
\end{table}

Table~\ref{tab:conversion} summarizes the outcome.
155 bundles---73 from the test device and 82 from firmware---now load and
predict through stock \coremltools{}, which is what makes the adversarial
campaign of Section~\ref{sec:adversarial} possible.
Three models resist conversion permanently and are informative in
themselves: \texttt{vmtracker} loads but fails inside the IR compiler at
prediction time (four per-platform copies), \texttt{MentionGenerationModel}
raises a compiler trap (three copies), and \texttt{hand\_gesture} segfaults
the compiler (four copies, Section~\ref{sec:crash}).

\subsection{Static analysis}
\label{sec:static_method}

The analyzer walks each artifact, identifies its kind
(\mlmodel{}, \mlmodelc{} directory, \mlmodelc{} archive, or loose \texttt{.dat}
blob), and extracts the specification version, model type, MIL op and dtype
histograms, weight-file inventory, and metadata provenance.
Three predicates drive Section~\ref{sec:weights}.
A bundle \emph{exposes plaintext parameters} if it is a protobuf whose
descriptor walk finds \texttt{WeightParams}, or whose model type stores
parameters inline (tree ensembles, text classifiers, MIL constants), or if
it is a compiled bundle with at least one weight blob that samples as raw
float32: from a $\le$64~KiB sample of every blob we unpack up to 4{,}096
fp32 values at seven header offsets and require $\ge$95\% finite values
with std $<10^6$ (an upper-bound sanity check---zero-variance samples
pass). Honesty note: 43 of the 281 raw-fp32 blobs (192.47 of
369.61~MiB, including 13 \texttt{megatron\_hsjs} wake-word blobs) rest
solely on zero-variance samples; under a stricter nonzero-variance
predicate the counts become 238 blobs (177.14~MiB).
A bundle is \emph{quantized} if it declares int8, 8-bit, fp16, or palettized
layers or ships fp16 blobs.
A bundle \emph{carries provenance} if its metadata contains an internal
build token, an author, a converter version, a version string, or a
user-defined dictionary.
The analyzer produced 1{,}137 records, one per artifact; every count in
Sections~\ref{sec:census}--\ref{sec:weights} is computed from this set.
The corpus boundary is the eight harvest sources of
Section~\ref{sec:harvesting} (six platform images and two device
harvests); the local-macOS \texttt{/System/Library} harvest is excluded
from it.

\subsection{Adversarial evaluation}

The adversarial campaign has two corpora that share one harness.
The first covers the models harvested directly (the firmware images): 56
models with classification, regression, embedding, image, audio, and
code-completion outputs, probed with 14 text-evasion generator families (8
direct techniques plus 6 multi-variant generators, e.g.\ homoglyph
substitution, zero-width insertion, case and punctuation manipulation,
edits, truncation, synonym substitution, and language mixing) spanning 32
distinct text attack types, and
a feature-space battery (dropout at 10\%, 30\%, and 50\%, Gaussian noise,
scaling by 0.1$\times$, 0.5$\times$, 2$\times$, and
10$\times$, extreme $\pm$50 perturbations, and all-zero, all-one, and
outlier injections).
The second applies the same battery to 165 models the conversion work made
available: the 155 conversion outputs of Section~\ref{sec:conversion} plus
10 already-loadable models shipped alongside them (SMS-filter, FinHealth,
and ghent-family classifiers, and the MCCKit BlackPearl family).
Token-level attacks target models that consume embeddings rather than text.
The two corpora total 82{,}092 campaign rows across 221 model
paths---82{,}085 \texttt{predict()} calls plus seven \texttt{\_\_population\_\_}
distribution-summary rows; the path count is over per-platform copies, not
distinct models. Every call runs in a subprocess with a timeout; 77{,}396
calls produced a comparable attacked output, and those records store
baseline and attacked outputs, the output distance, and the flip flag.
Section~\ref{sec:adversarial} accounts for the calls without a scored
output---4{,}680 \texttt{crema}/TFModel battery calls, four failed runs, and
five clean-repeat sanity rows.

\subsection{On-device latency and poisoning harnesses}

Latency is measured on the test device (the virtual instance of
Section~\ref{sec:harvesting}), not on a workstation: a
benchmark app loads each system model through
\texttt{MLModel(contentsOf:)} with \texttt{MLComputeUnitsAll} and reports
the median of 50 predictions per condition, with cold and warm load times
measured separately.
A Frida hook on the prediction path records the per-stage breakdown.
The poisoning harness modifies weight files in place under
\texttt{/System/Library}, restores them from pristine copies afterwards, and
verifies restoration by SHA-256.

\section{Results}
\label{sec:results}

\subsection{Census: what ships}
\label{sec:census}

\begin{figure*}[t]
\centering
\includegraphics[width=\textwidth]{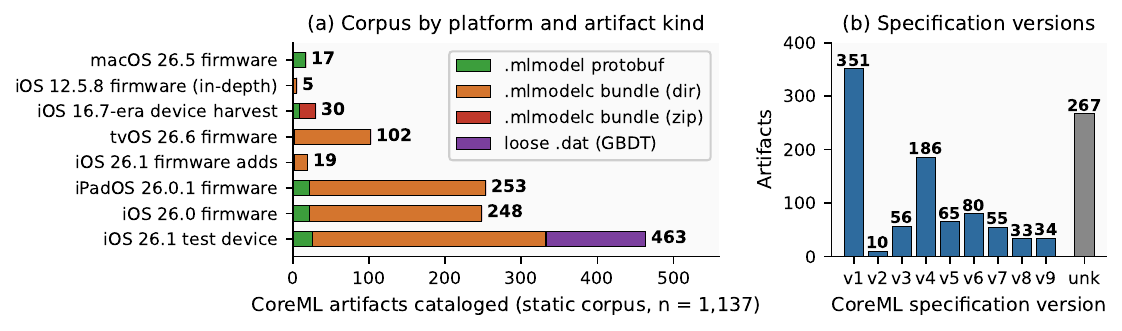}
\caption{The static-analysis corpus. (a)~1{,}137 CoreML artifacts by
platform and artifact kind: the iOS~26.1 test device contributes 463
artifacts; the iOS~26.0 and iPadOS~26.0.1 firmware images 248 and 253; a
firmware-extracted iOS~26.1 add-on set 19; tvOS~26.6 102; an iPhone~X
iOS~16.7-era device harvest 30; the iOS~12.5.8 firmware 5 (analyzed in
depth); and a macOS~26.5 firmware image 17.
(b)~Specification versions of the same corpus: versions~1 through~9 are all
present in shipped artifacts, and 267 records carry no parseable
specification version (loose \texttt{.dat} tree-ensemble blobs and
non-protobuf inputs).}
\label{fig:census}
\end{figure*}

On the iOS~26.1 test device we inventory \textbf{320 model files across 54
framework groups}---one entry per model file, grouped by family: 53 named
families plus one unattributed group of 54 models.
The largest families are voice services (32 models: text-to-speech
encoders, decoders, vocoders for four locales), Spotlight ranking (27),
power management (22), sound analysis (17), text recognition (15),
preinstalled assets (14), people suggestions (10), health (8),
intelligence platform (7), and SMS filtering, finance, scene understanding,
mail categorization, and Siri inference (6 each); media analysis
contributes 4, including the hand-gesture model of Section~\ref{sec:crash},
and wireless resource management 3.
(The record set of Figure~\ref{fig:census} counts 463 artifacts on the same
device, adding loose blobs and per-asset duplicates the inventory's
one-entry-per-file rule excludes.)

The static corpus of Figure~\ref{fig:census}a includes firmware and macOS
and counts artifact kinds separately: 1{,}137 artifacts total---98
\mlmodel{} protobufs, 887 \mlmodelc{} directories, 22 extracted \mlmodelc{}
archives (21 from the iPhone~X harvest), and 130 loose weight blobs (ATX
feature stores, named-entity dictionaries, Vision face-model blobs,
tree-ensemble \texttt{\_B0000.DAT} files, and media-analysis
\texttt{cnn\_*.dat} arrays).
The per-platform split is in Figure~\ref{fig:census}a.

\paragraph{Cross-platform byte identity, with limits.}
All 22 iPadOS~26.0.1 protobufs are SHA-256-identical to their iOS~26.0
counterparts: in the cases we compared, Apple compiles one artifact per
model and redistributes the same bytes per platform.
The sharing extends to tvOS: 84 of its 100 bundles have a byte- or
metadata-identical counterpart in the other firmware harvests (51
byte-identical, 33 metadata-identical), both tvOS \mlmodel{} protobufs
are SHA-256-identical to their iOS copies, and 10 bundles are TV-exclusive
(color calibration, audio analysis, app ranking).
The remaining 6 name-shared bundles match neither at either level; the
artifact package carries the per-bundle drift table (listed in the Open
Science manifest), and the cases are
small and specific: three differ only in Apple's embedded internal
\texttt{mldb} build token, with byte-identical weight files
(\texttt{HPdetector\_}\allowbreak\texttt{SceneClassifier\_}\allowbreak\texttt{B238}/\texttt{B520},
\texttt{megatron\_}\allowbreak\texttt{hsjs\_}\allowbreak\texttt{8bit\_}\allowbreak\texttt{compressed\_}\allowbreak\texttt{v1});
\texttt{PlanResolutionModel} ships a newer Float32 build on tvOS
(v9.2.0 vs.\ Float16 v9.0.0); \texttt{social\_}\allowbreak\texttt{highlights\_}\allowbreak\texttt{scorer}
is a different single-input model (generated class
\texttt{social\_}\allowbreak\texttt{highlights\_}\allowbreak\texttt{tv\_}\allowbreak\texttt{scorer}); and
\texttt{anst}'s tvOS copy omits the \texttt{model.espresso.*} graph and the
per-device precompilation blobs.
The SMS-filter copies match as well: the test device's framework copy of
\texttt{mainModel\_}\allowbreak\texttt{INEN} is SHA-256-identical to the iOS and iPadOS firmware
copies, and its deterministic attack families (zero-width insertion)
produce bit-identical attacked outputs across all three
(Section~\ref{sec:adversarial}); the randomized text-corruption families
draw fresh randomness per run and differ in 11--16 of 1{,}422 records per
copy-pair.
A defect or a poisoning payload therefore transfers across every platform
sharing the artifact's bytes.

\paragraph{Legacy artifacts stay deployed.}
The iOS~12.5.8 image still ships five \mlmodelc{} bundles (ATX notification
suggestions, wrist detection, a GAN, and two Spotlight rankers) whose stored
weights are unencrypted in every container our analyzer recognizes (two
ship sampled raw-fp32 blobs, two gradient-boosted decision tree (GBDT)
files, one carries no inventoried blob) and whose metadata predates the modern toolchain; the
\texttt{spotlight\_l2} ranker persists on iOS~26.x in re-trained form;
\texttt{spotlight\_l2\_recency}---whose role Spotlight now folds into
\texttt{spotlight\_l3\_ios}---and wrist detection have been retired,
not dead code.

\begin{figure*}[t]
\centering
\includegraphics[width=\textwidth]{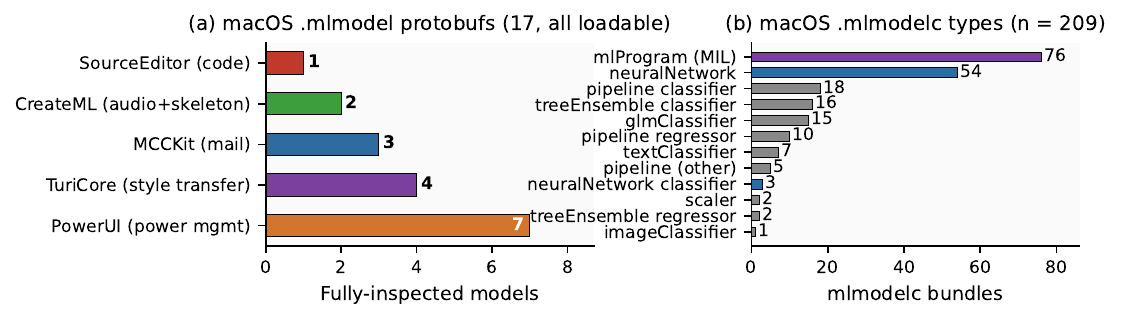}
\caption{The macOS on-device model landscape. (a)~The 17 fully inspected
\mlmodel{} protobufs by framework family. (b)~Model-type distribution of
the 209 \mlmodelc{} bundles with parseable \texttt{metadata.json} from the
local \texttt{/System/Library} harvest, outside the 1{,}137-artifact
corpus; \mlprogram{} and \texttt{neuralNetwork} together account for 62\%.}
\label{fig:macos}
\end{figure*}

\paragraph{The macOS subset.}
The macOS~26.5 image contributes the corpus's 17 \mlmodel{} protobufs: all
17 load, 14 predict, and three BlackPearl \mlprogram{} models need an
external weights blob the extraction omits (Figure~\ref{fig:macos}).
They cover battery-charge prediction (\texttt{PowerUI} tree ensembles with
840--5{,}534 nodes and up to 407 features), Turi Create-style transfer
models~\cite{turicreate} (a 29.1~MiB VGG-16 among them), the BlackPearl
mail-categorization models at specification version~6, CreateML audio
embeddings plus a CreateMLComponents skeleton-action model, and an
18.2~MiB recurrent code-completion model from
MobileAsset.
A separate harvest of a local (non-image) macOS~26.5
\texttt{/System/Library} found 361 \mlmodelc{} directories (209 with
parseable \texttt{metadata.json}) and 102 AssetsV2 bundles;
Figure~\ref{fig:macos}b shows the model-type distribution of the 209
parseable bundles. This harvest stays outside the 1{,}137-artifact corpus,
which counts only the 17 macOS protobufs.

\begin{figure}[t]
\centering
\includegraphics[width=\columnwidth]{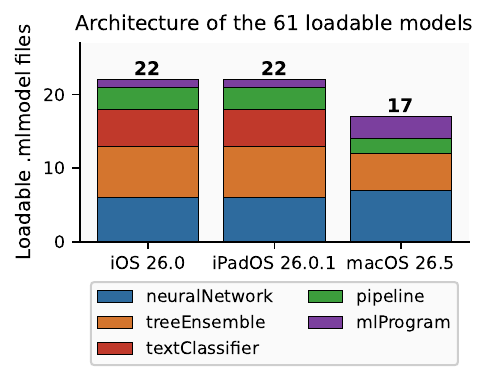}
\caption{Architecture landscape of the 61 directly loadable \mlmodel{}
protobufs from the initial firmware and macOS harvest, before conversion.
iOS~26.0 and iPadOS~26.0.1 carry the identical 22-model set; macOS
contributes 17. Conversion later rendered 155 additional bundles
loadable (82 firmware, 73 test-device).}
\label{fig:arch_landscape}
\end{figure}

Figure~\ref{fig:arch_landscape} shows the architecture mix of the
protobuf-loadable set; the 155 converted bundles extend it with models
Apple shipped only in compiled form.

\subsection{Format evolution and locale divergence}
\label{sec:formats}

\paragraph{Nine specification generations in production.}
The specification-version histogram of Figure~\ref{fig:census}b spans
v1 (351 artifacts), v2 (10), v3 (56), v4 (186), v5 (65), v6 (80), v7 (55),
v8 (33), and v9 (34). Version~1 artifacts include the oldest GBDT bundles
and legacy GLM/SVM classifiers; v6 and above are mostly MIL-based
\mlprogram{}
models, with 15 v8 CreateML \texttt{textClassifier} counterexamples.
The version is not a compatibility boundary: a single device hosts models
two generations apart, so the toolchain must span the range.

\paragraph{Four weight containers.}
Independent of the specification version, learned parameters appear in four
container formats (Table~\ref{tab:formats}).
All four are raw numeric data with at most a small index header.

\begin{table}[t]
\centering
\caption{Weight container formats found in the corpus.
Most formats store weights as raw fp32/fp16/int8 values: for 284 of the
687 inventoried blobs (281 raw-fp32, 3 fp16), a 64~KiB window scan
confirms raw float distributions, and a further 37 sub-64-byte blobs are
too small to type-check; the remaining 366 (including all 92
\texttt{\_B0000.DAT} GBDT blob entries, 91 distinct files) are opaque to
the scan and were read with format-specific parsers.}
\label{tab:formats}
\scriptsize
\setlength{\tabcolsep}{4pt}
\begin{tabular}{@{}>{\raggedright\arraybackslash}p{3.1cm}>{\raggedright\arraybackslash}p{3.4cm}@{}}
\toprule
\textbf{Format} & \textbf{Found in} \\
\midrule
\texttt{model.}\allowbreak\texttt{espresso.}\allowbreak\texttt{weights} & espresso neural-network bundles; small uint32 index header, then weight values \\
\texttt{weights/weight.bin} & MIL bundles, spec $\geq$6; header then raw fp32 (fp16/int8 when quantized) \\
\texttt{\_B0000.DAT}, \texttt{*.dat} & tree-ensemble bundles and loose blob arrays; raw fp32 node/leaf weights, no header \\
\texttt{.mlmodel} \texttt{WeightParams} & protobuf models; inline fp32, fp16, or raw values; quantized layers via the \texttt{quantization} field \\
\bottomrule
\end{tabular}
\end{table}

\paragraph{Quantization is widespread but partial.}
262 of 1{,}137 artifacts declare quantized storage---int8, 8-bit, fp16, or
palettized. A related shipped case (quantized by declaration---its name
and metadata say 8-bit compressed---but not among the 262 under our strict
storage predicate, which keys on declared layer types and fp16 blobs) is
the Siri wake-word transformer family recovered from the iOS~26.1 image
extract (Section~\ref{sec:harvesting}): 14 instances of a five-layer
encoder with a 163-class output head (9 \texttt{hsjs}-variant, 5 plain),
shipped 8-bit compressed; the \texttt{hsjs} variant carries 42 dynamic
per-layer scale factors. Apple
has published high-level descriptions of the voice-trigger
system~\cite{siri_voice_trigger}; our recovery is of the deployed
artifact---its exact layer map, quantization scheme, and internal codename.
Quantization reduces file size but not readability: an fp16 or int8 blob is
as directly interpretable as an fp32 one.

\begin{figure}[t]
\centering
\includegraphics[width=\columnwidth]{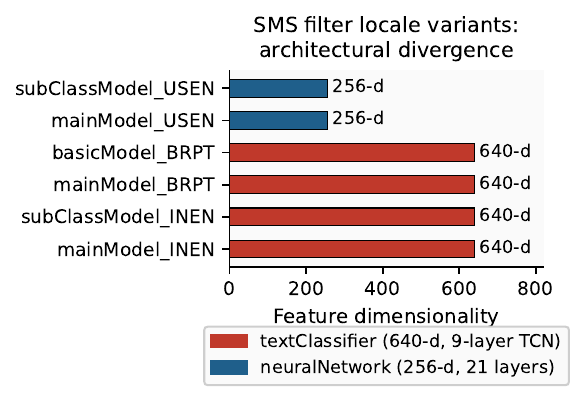}
\caption{Architectural divergence inside the SMS filter family. BRPT and
INEN variants are Create~ML \texttt{textClassifier} pipelines (string
input, 640-dimensional embedding, 9-layer TCN); the USEN variants are raw
21-layer \texttt{neuralNetwork} models with a tokenized multiArray input.}
\label{fig:locale_arch}
\end{figure}

\begin{figure}[t]
\centering
\includegraphics[width=\columnwidth]{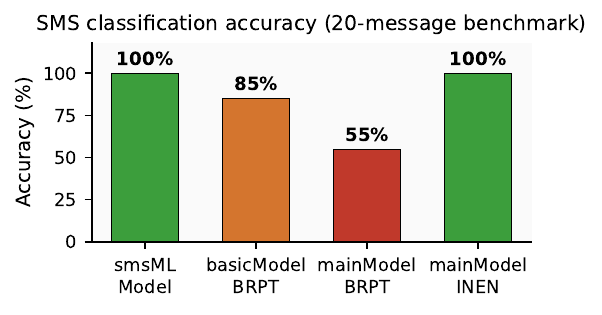}
\caption{SMS classification accuracy on the 20-message benchmark.
The USEN variants use a
different input contract and are evaluated adversarially rather than on
text accuracy (Section~\ref{sec:adversarial}).}
\label{fig:accuracy}
\end{figure}

\begin{figure}[t]
\centering
\includegraphics[width=\columnwidth]{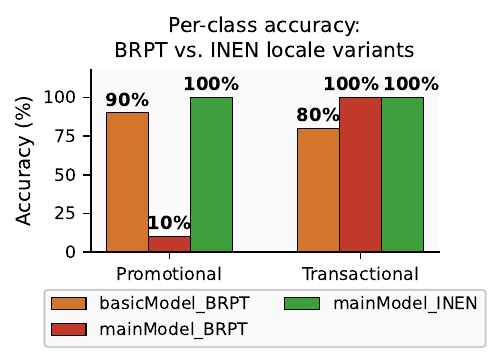}
\caption{Per-class accuracy of the locale variants on the 20-message SMS
benchmark. \texttt{mainModel\_}\allowbreak\texttt{BRPT} labels only 10\% of promotional
messages correctly while \texttt{mainModel\_}\allowbreak\texttt{INEN} is perfect on both
classes.}
\label{fig:locale}
\end{figure}

\paragraph{Locale variants diverge architecturally, not just in weights.}
The SMS filter ships as separate model sets per locale, and the divergence
is structural.
The Brazilian-Portuguese (BRPT) and Indian-English (INEN) variants are
Create~ML text classifiers: a 9-layer temporal convolutional network (TCN)
with a 640-dimensional text embedding, taking raw text as input.
A field-by-field comparison of the descriptors shows a single differing
embedding selector---\texttt{RptWDynamic} in BRPT, \texttt{RenWDynamic} in
INEN---associated with a 45-point accuracy gap on our 20-message benchmark:
\texttt{mainModel\_}\allowbreak\texttt{INEN} and the iOS~16.7-era \texttt{smsMLModel} score 100\%,
\texttt{basicModel\_}\allowbreak\texttt{BRPT} 85\%, and \texttt{mainModel\_}\allowbreak\texttt{BRPT} 55\%
(Figures~\ref{fig:accuracy},~\ref{fig:locale}).
The US-English (USEN) variants abandon that pipeline: raw 21-layer networks
converted from PyTorch~2.1.2 through \coremltools~7.1.4, consuming a
tokenized \texttt{multiArray[1,512,1,256]} input rather than text
(Figure~\ref{fig:locale_arch}).
A user's locale thus determines not only which weights run but which model
family, input contract, and failure mode they get.

Table~\ref{tab:sms} summarizes the family as harvested: five Create~ML
string-input classifiers and the two raw-network USEN
variants.

\begin{table}[t]
\centering
\caption{The shipped SMS-filter family and its variants. Benchmark
accuracy is on the 20-message set, measured through the \coremltools{}
macOS runtime---the only runtime where all five Create~ML classifiers
load (in our initial test-device session, \texttt{mainModel\_}\allowbreak\texttt{BRPT} failed
to load for lack of its dynamic-embedding MobileAsset; a later session on
the same device, after the asset arrived, loaded it and produced
predictions). Latency for these
models is reported separately as on-device medians
(Section~\ref{sec:latency}).}
\label{tab:sms}
\scriptsize
\setlength{\tabcolsep}{4pt}
\begin{tabular}{@{}>{\raggedright\arraybackslash}p{2.25cm}>{\raggedright\arraybackslash}p{0.85cm}>{\raggedright\arraybackslash}p{2.55cm}r@{}}
\toprule
\textbf{Model} & \textbf{Locale} & \textbf{Architecture} & \textbf{Acc.} \\
\midrule
\texttt{mainModel\_}\allowbreak\texttt{INEN} & INEN & 9-layer TCN, \texttt{RenWDynamic}, 1.42~MiB & 100\% \\
\texttt{mainModel\_}\allowbreak\texttt{BRPT} & BRPT & 9-layer TCN, \texttt{RptWDynamic} & 55\% \\
\texttt{basicModel\_}\allowbreak\texttt{BRPT} & BRPT & older (unreconstructed), \texttt{QWDynamic}, spec v3 & 85\% \\
\texttt{subClassModel\_}\allowbreak\texttt{INEN} & INEN & TCN, own 4-class taxonomy & --- \\
\texttt{smsMLModel} & --- & TCN (iPhone~X, 16.7 era) & 100\% \\
\texttt{mainModel\_}\allowbreak\texttt{USEN} & USEN & 21-layer NN (PyTorch 2.1.2) & --- \\
\texttt{subClassModel\_}\allowbreak\texttt{USEN} & USEN & 21-layer NN, tokenized input & --- \\
\bottomrule
\end{tabular}
\end{table}

The two-tier design is visible in the naming and the shipped thresholds:
tier~1 (\texttt{mainModel}) decides promotional versus transactional;
tier~2 (\texttt{subClassModel}) routes transactional messages into Orders,
Reminders, Finance, or Others; a separate FinHealthCore \texttt{intent\_classification} model (transaction\_SQL vs. nontransaction\_SUPPORT) addresses transaction
sub-types. Because tier~2 consumes the same text through the same
tokenization path, it is not an independent check: the campaign shows the
two decisions failing separately
(Section~\ref{sec:adversarial}). A tier-1 pass still guarantees
nothing, because the second
decision is neither independent nor strictly narrower; defense-in-depth
would require the tiers to disagree by construction---different inputs,
normalization, or an independent signal---and as shipped they do not.

\paragraph{Thresholds are shipped with the models.}
Alongside the USEN filter models, the test device carries the production
inference thresholds in \texttt{trialThreshold\_USEN.plist}: a Promotional
threshold of 0.85 and subcategory thresholds of 0, with empty regular
expression lists.
The pipeline is therefore model-only: any message scoring below 0.85 on the
promotional head falls through to the transactional class with no regex or
rule-based safety net, so each adversarial flip in
Section~\ref{sec:adversarial} converts directly into a transactional label;
within the released model pipeline, no second stage inspects what the text
filter passes.

\subsection{Adversarial robustness}
\label{sec:adversarial}

\begin{figure}[t]
\centering
\includegraphics[width=\columnwidth]{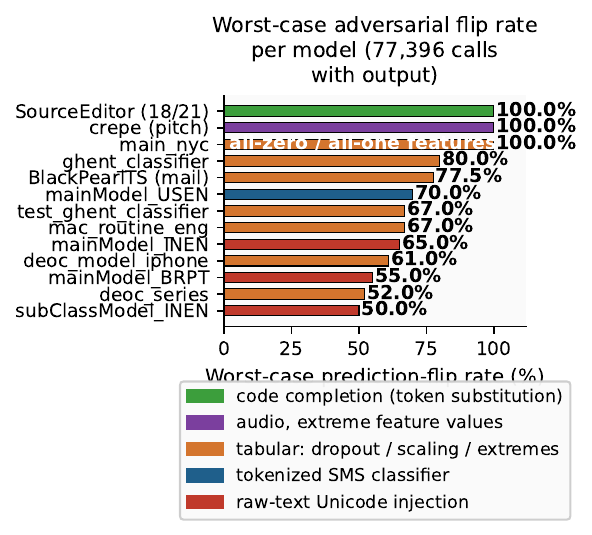}
\caption{Worst-case prediction-flip rate for a representative selection of
13 models from the combined campaign (221 model paths in total: 82{,}092
campaign rows, of which 77{,}396 \texttt{predict()} calls carry an attacked
output). Bars are colored by attack family and show each model's worst case
over all attacks measured; the two annotated cases call out their mechanism
(\texttt{main\_nyc} flips under constant-feature probes; \texttt{SourceEditor}
is the 18-of-21 token-substitution cross-check of Table~\ref{tab:adv};
its 100\% bar rests on single-probe groups). Not
shown: \texttt{BlackPearl} at 82.5\% (33/40), above every plotted bar except
the three 100\% models (SourceEditor, crepe, and main\_nyc). The \texttt{mainModel\_}\allowbreak\texttt{USEN} bar, like every USEN
adversarial number here, measures the shipped network under the surrogate
embedding reconstruction of Section~\ref{sec:adversarial}.}
\label{fig:adv_flips}
\end{figure}

The combined campaign covers 221 model paths: 56 harvested directly and
165 in the converted-corpus run (155 conversion outputs plus 10
already-loadable models). Its 82{,}092 rows comprise 82{,}085
\texttt{predict()} calls plus seven \texttt{\_\_population\_\_} summary
rows---one per tree-regressor path, recording the 20-input prediction
distribution used to normalize the regressor shifts (they are not calls).
Of the calls, 77{,}396 produced a comparable attacked output; four further
runs failed at input validation (a TTS encoder,
\texttt{fastspeech2\_encoder}, rejecting all-zero tensor inputs---correct
behavior, not a harness failure) and are counted among the non-output
rows. Of the 4{,}685
completed calls that carry no attacked output, 4{,}680 are \texttt{crema}
and \texttt{TFModel} battery calls with no recorded model output (the
harness captured neither a baseline nor an attacked response; they
contribute latency only) and five are clean-repeat sanity rows. No
execution crashed or hung.
Figure~\ref{fig:adv_flips} shows the worst case per model;
Table~\ref{tab:adv} collects the same by category.

The two runs were executed about four weeks apart (2026-08-01 and
2026-09-01) with the same attack battery, which is why deterministic
results for shared models reproduce across platforms and runs.
Run~1 (56 models) contributes 47{,}371: 14{,}220 raw-text and
3{,}208 tokenized-USEN text-classifier probes, 22{,}000 tree-ensemble
classifier and 7{,}700 regressor perturbations (plus seven
\texttt{\_\_population\_\_} summary rows, one per regressor path), 236
network/audio/special. Run~2 (165 models) contributes
34{,}721: 8{,}074 classifier, 18{,}200 regressor, 7{,}110 string-input
text-classifier, 1{,}025 special (4 rejected), 240 embedding, 72 audio.

\paragraph{Unicode attacks defeat the best text classifiers.}
Injecting zero-width characters (\zwsp{}, \zwj{}, \zwnj{}) between
characters of an SMS flips \textbf{65\%} of \texttt{mainModel\_}\allowbreak\texttt{INEN}
predictions (13 of 20 benchmark messages)---the model with 100\% baseline
accuracy---identically on all three recorded copies (iOS, iPadOS, and the
test device's framework copy), whose checksums match and whose attacked
outputs are bit-identical (Section~\ref{sec:formats}).
Replacing Latin characters with Cyrillic homoglyphs at 100\% rate flips
50\% of both \texttt{mainModel\_}\allowbreak\texttt{INEN} and \texttt{subClassModel\_}\allowbreak\texttt{INEN}, and
30\% substitution flips 10--35\% depending on the model and platform copy
(\texttt{subClassModel\_}\allowbreak\texttt{INEN} is the more sensitive of the two).
The tokenized USEN variant is not immune, but it is measured under a
surrogate pipeline. The production tokenizer and embedding are not shipped
in any artifact we recovered---the network consumes a
$1{\times}512{\times}1{\times}256$ tensor whose upstream
text$\rightarrow$embedding stage remains unresolved---so every USEN
adversarial number in this paper measures the \emph{shipped network} under
a reconstruction: a deterministic per-character pseudo-random
256-dimensional embedding (seeded by character code) over a 512-character
window. Under that surrogate, token-level \zwsp{} insertion flips
\textbf{70\%} of \texttt{mainModel\_}\allowbreak\texttt{USEN} predictions (14 of 20); we report
it as an approximate upper bound for the deployed pipeline, not a
production-model robustness number.

\paragraph{Feature-space attacks defeat tabular models.}
Dropping 50\% of input features flips 67\% of \texttt{test\_ghent\_classifier}
(an income-category model) and 61\% of \texttt{deoc\_model\_iphone};
10$\times$ feature scaling flips 67\% of \texttt{mac\_routine\_eng} and 52\%
of \texttt{deoc\_series}; the BlackPearl mail-classification ensemble
(\texttt{BlackPearlTS}) flips 77.5\% of its probes under the extreme $-50$
perturbation (Figure~\ref{fig:adv_flips}).
The PowerUI charge-prediction ensembles (\texttt{deoc\_*}) are uniformly
fragile: their least sensitive member (\texttt{deoc\_ultra}) still flips
36\% under the same battery, and even the five-feature \texttt{shallow}
proxy flips 19\%.
Regression variants shift proportionally: $0.1\times$ scaling moves
\texttt{dynamic\_scheduling} by a mean relative shift of 0.75 (mean
$|\Delta\text{pred}|\approx 13.1$, about five times the standard deviation
of the model's own predictions over the 20 recorded inputs
($\sigma=2.59$), and \texttt{EntityReranker} shifts its output by a mean of
49{,}936\% under an extreme perturbation (worst single probe
$\times$6{,}908).
Extreme-value attacks alone reach 100\% flip rates on the pitch model
(\texttt{crepe}) and 80\% on \texttt{ghent\_classifier}; the all-zero and
all-one feature attacks reach 100\% on an unidentified shipped
\mlprogram{} (\texttt{main\_nyc}) (90\% under extreme $\pm$50).

\paragraph{Model-quality defects surface under attack.}
A shipped style-transfer model (\texttt{resnet-16}) produces a strong
style-0 stylization of neutral-gray inputs that is invariant to
brightness perturbations (identical output L2 across attacks, with
substantial deviation from the gray source; the measurement is released as
two supplementary dataset rows, Appendix~\ref{app:open})---an output that
does not degrade gracefully.
An app-item network named for its explicit feature ordering is
\emph{not} column-order invariant: permuting its 471 feature columns changes
the output by L2~${\approx}~15.2$--$16.4$, the same magnitude as shuffling
rows (15.6--15.8).
A bandwidth-prediction LSTM (long short-term memory network) ignores its
state inputs entirely (zeroing them changes nothing, distance 0.000) and
responds to input noise with shifts of 0.084--0.139 (mean
${\approx}0.11$), consistent with a stub rather than the intended streaming
model.
The code-completion model flips its top-1 token on 18 of 21 single-token
substitutions measured in our campaign (86\%; flipping-substitution logit
L2 shifts 16--86, e.g.\ 38.9 at token 500) and on all six sequence-length
changes (L2 94--120), but we flag the result as
construction-dependent: an independent cross-check with a different input
construction observed the same top-3 tokens, no top-1 flip, and an L2 shift
of only 4.5. We report the discrepancy unresolved.

\begin{table}[t]
\centering
\caption{Worst-case adversarial result per model category across the
82{,}092-row campaign (77{,}396 calls with an attacked output).}
\label{tab:adv}
\scriptsize
\setlength{\tabcolsep}{4pt}
\begin{tabular}{@{}>{\raggedright\arraybackslash}p{2.4cm}>{\raggedright\arraybackslash}p{2.9cm}>{\raggedright\arraybackslash}p{2.3cm}@{}}
\toprule
\textbf{Category} & \textbf{Worst-case attack} & \textbf{Effect} \\
\midrule
SMS text classifier & \zwsp{}/\zwj{}/\zwnj{} insertion, \texttt{mainModel\_}\allowbreak\texttt{INEN} & 65\% flip (100\%$\rightarrow$35\%) \\
SMS classifier (tokenized) & token-level \zwsp{}, \texttt{mainModel\_}\allowbreak\texttt{USEN} & 70\% flip (surrogate embedding) \\
SMS text classifier & homoglyph substitution, \texttt{subClassModel\_}\allowbreak\texttt{INEN} & 50\% flip \\
Tree ensemble (classifier) & 50\% dropout, \texttt{test\_ghent\_classifier} & 67\% flip \\
Tree ensemble (classifier) & 10$\times$ scaling, \texttt{mac\_routine\_eng} & 67\% flip \\
Tree ensemble (regressor) & $0.1\times$ scaling, \texttt{dynamic\_scheduling} & 0.75 mean rel.\ shift ($\approx$5.0 std, mean $|\Delta|$ 13.1) \\
Pitch model (\texttt{crepe}) & extreme $-50$ perturbation & 100\% flip \\
Unidentified \texttt{mlProgram} (\texttt{main\_nyc}) & all-zero / all-one features & 100\% flip \\
Neural network (image) & brightness shift, \texttt{resnet-16} style 0 & output invariant to perturbation \\
LSTM (bandwidth) & $\sigma{=}1.0$ noise & 0.084--0.139 shift; state ignored \\
Feature-order network & column permutation & L2 {$\approx$} 15.2--16.4 \\
Code completion & single-token substitution & top-1 flip, 18/21 (construction-dependent) \\
\bottomrule
\end{tabular}
\end{table}

\paragraph{Phishing passes the filter at baseline.}
Before any adversarial manipulation, the three iOS SMS classifier families
label \textbf{8--9 of 10} hand-crafted phishing messages (80--90\%,
$n=10$) as Transactional---the model output that drives the filter's
decision. (An earlier four-message probe reported 75--100\%; the
ten-message set does not confirm the upper end.)
Live device probes agree: a lure-style message is labeled
Transactional by \texttt{mainModel\_}\allowbreak\texttt{INEN} and \texttt{Others} by
\texttt{subClassModel\_}\allowbreak\texttt{INEN}.
These are classifier labels, not observed inbox outcomes; end-to-end inbox
routing is not demonstrated.
The filter is a promotional-spam classifier in practice, not a phishing
one, and the two-tier chain offers no second chance: under \zwsp{}
injection tier~1 flips 13 of 20 messages and tier~2 flips 10, with both
tiers flipping together on only 5, and on 5 further messages tier~2 flips a
message tier~1 left alone---so most tier~1 passes do not survive tier~2
(5 of the 7 tier~1 passes flip at tier~2; the remaining 2 survive on all
three recorded copies).

\paragraph{A systematic class bias.}
Across the 4{,}266 baseline predictions recorded for this model in the
campaign (4{,}053 ``Transactional'' and 213 ``Promotional'', identically
1{,}351/71 on each of the three copies), \texttt{mainModel\_}\allowbreak\texttt{BRPT} emits a
95:5 label skew that matches its 10\% promotional accuracy and is
consistent with the collapsed-embedding hypothesis of
Section~\ref{sec:formats} rather than a benchmark artifact.

\subsection{Weight security: plaintext by construction}
\label{sec:weights}

\begin{table}[t]
\centering
\caption{Static-analysis results over 1{,}137 artifacts
(the static analyzer of Section~\ref{sec:static_method}).}
\label{tab:static}
\scriptsize
\setlength{\tabcolsep}{4pt}
\begin{tabular}{@{}lr@{}}
\toprule
\textbf{Metric} & \textbf{Count} \\
\midrule
Artifacts analyzed & 1{,}137 \\
Artifacts exposing plaintext parameters & \textbf{368} \\
Weight blobs inventoried & 687 (762.62~MiB) \\
Quantized bundles (fp16/int8/8-bit) & 262 \\
Bundles exposing provenance metadata & 463 \\
\quad of which carry an \texttt{mldb\_token} & 166 \\
Bundles with \texttt{isUpdatable=1} & 11 \\
\bottomrule
\end{tabular}
\end{table}

Table~\ref{tab:static} collects the analyzer's headline counts.
Of the 1{,}137 artifacts, \textbf{368 expose learned parameters in
plaintext} (281 blobs, 369.61~MiB); corpus-wide, our inventory records
687 weight blobs (762.62~MiB). None of these containers carries
encryption or obfuscation our analyzer recognizes, and---per our
64~KiB sampling scan---no key material appears in the sampled regions.
The four container formats of Table~\ref{tab:formats} share one property: a
scanner can unpack a 64~KiB sample, confirm raw float distributions, and
extract every stored parameter---there is no format requiring an attacker
to do more than read a file.

\paragraph{Provenance leaks the build pipeline.}
463 artifacts carry provenance metadata, and 166 of 166 MIL bundles---every
bundle containing a \texttt{model.mil}---carry Apple's internal
model-database build token (\texttt{mldb-}\allowbreak\texttt{XXXXXXXXXX}), compiler component
version (\texttt{coremlc-}\allowbreak\texttt{component-}\allowbreak\texttt{MILHost}), exporter version, and
training-stack strings such as
\texttt{coremltools-}\allowbreak\texttt{component-}\allowbreak\texttt{tensorflow:~2.11.0}.
The tokens are unique per build, so they correlate releases and fingerprint
unannounced models; one description also contains an internal \texttt{s3://}
asset path (tested against public S3 endpoints, which return 404): the
identifier leaks naming structure, not a reachable bucket.

\subsection{On-device poisoning: 16 bytes, no verification}
\label{sec:poisoning}

\begin{table}[t]
\centering
\caption{In-place weight poisoning on the iOS~26.1 test device. All
modifications were restored and verified byte-identical afterwards.}
\label{tab:poison}
\scriptsize
\setlength{\tabcolsep}{4pt}
\begin{tabular}{@{}>{\raggedright\arraybackslash}p{2.4cm}>{\raggedright\arraybackslash}p{4.6cm}@{}}
\toprule
\textbf{Edit} & \textbf{Effect} \\
\midrule
16~B in the interactionPrediction weight file & probes flip \texttt{clustering}$\rightarrow$\texttt{recency} at $p{=}1.0$ (7/8 inputs; the eighth becomes a 50/50 tie) \\
64~B sign flip in the ATXWidgetPredictionMLModel final layer & all probes flip \texttt{Not Engaged}$\rightarrow$\texttt{Engaged} \\
0xFF fill of the \texttt{update}\allowbreak\texttt{Parameters/} metadata & no effect: that directory is not read at inference \\
\bottomrule
\end{tabular}
\end{table}

We poisoned two system models in place, loaded through their exact
\texttt{/System/Library} URLs, on the test device (root, jailbroken---the
A2b class of Section~\ref{sec:threat}).

\paragraph{16 bytes flip 7 of 8 probe predictions.}
Overwriting the four dense weights at offset 0xc0 of
\texttt{interaction}\allowbreak\texttt{Prediction}'s 208-byte \texttt{model.}\allowbreak\texttt{espresso.}\allowbreak\texttt{weights}
with saturating values flips the model's output from
\texttt{clustering} (``group these events'') to \texttt{recency}
(``show the most recent'') at $p=1.0$ on seven of eight probe
inputs; the eighth becomes a degenerate 50/50 tie.
An analogous 64-byte sign flip of the final classification layer in
\texttt{ATXWidget}\allowbreak\texttt{Prediction}\allowbreak\texttt{MLModel} flips every probe from
\texttt{Not Engaged} to \texttt{Engaged} (Table~\ref{tab:poison}).
Both models feed suggestion surfaces, so poisoned behavior is a product
decision, not a numeric glitch.
The demonstration is at the model-API boundary: we loaded the consumers'
exact bundles and observed flipped predictions, but did not observe a
consuming daemon issue one in situ.

\paragraph{Zero runtime integrity verification.}
In the code paths we inspected, nothing verifies the weight bytes---no
checksum, no signature on data files, no CoreML-side seal check, and no
trustcache entry for model data; arbitrary valid-format edits load and
execute.
The only control observed is the sealed system volume.

\paragraph{The seal is the only barrier, and it is one boot operation
deep.}
\texttt{mount -uw /} makes the sealed root volume writable on a jailbroken
device; \texttt{mount -ur /} then fails with \texttt{errno}~66, so the
volume stays writable for the rest of the boot.
A reboot re-seals the volume and restores baseline predictions (verified),
making runtime-only root a non-persistent attacker.
A runtime-root attacker on a stock device (A2a) can read every model and
write non-sealed paths but cannot perform this step.
An attacker with boot-chain control (A2b), however, owns the seal itself,
and since we found no per-file verification afterwards, poisoned models
can persist with the seal intact.
All modified files were restored and SHA-256-verified before and after the
reboot.

\paragraph{The updatable path is not the inference path.}
Corrupting the \texttt{update}\allowbreak\texttt{Parameters/} directories of the same bundles
(45- and 71-byte \texttt{coremldata.bin} payloads listing the trainable
layers \texttt{dense\_1} and \texttt{dense\_concatenated}) produces output
identical to pristine.
These directories are metadata for \texttt{MLUpdateTask}; inference reads
the shipped weight files directly, so an attacker edits weights, not
training metadata---and the platform verifies neither.

\subsection{The updatable-model surface}
\label{sec:updatables}

\begin{table}[t]
\centering
\caption{Bundles shipped with \texttt{isUpdatable=1} across the corpus.
Each row is a distinct model; the platforms column lists every corpus copy
that carries the flag.}
\label{tab:update}
\scriptsize
\setlength{\tabcolsep}{4pt}
\begin{tabular}{@{}>{\raggedright\arraybackslash}p{3.9cm}>{\raggedright\arraybackslash}p{4.0cm}@{}}
\toprule
\textbf{Model} & \textbf{Copies with \texttt{isUpdatable}} \\
\midrule
\texttt{interaction}\allowbreak\texttt{Prediction} & iOS 26.0, iPadOS 26.0.1, tvOS 26.6, 26.1 test device \\
\texttt{HLSML}\allowbreak\texttt{Bandwidth}\allowbreak\texttt{Predictor} & iOS 26.0, iPadOS 26.0.1, tvOS 26.6, 26.1 test device \\
\texttt{ATXWidget}\allowbreak\texttt{Prediction}\allowbreak\texttt{MLModel} & iOS 26.0, iPadOS 26.0.1, 26.1 test device \\
\bottomrule
\end{tabular}
\end{table}

Eleven artifacts carry \texttt{isUpdatable=1} (Table~\ref{tab:update}):
three model families---interaction prediction, streaming-bandwidth
prediction, and ATX widget engagement---each replicated across three or
four platform copies. Two security consequences follow. First, the bundle
ships an \texttt{update}\allowbreak\texttt{Parameters/} directory describing the trainable layers
through \texttt{MLUpdateTask}~\cite{mlupdatetask}; the shipped copies are
never written by the update flow (Section~\ref{sec:poisoning}), but the
mechanism shows CoreML's design assumes weight files can be replaced at
runtime by a training daemon. Second, the selection is itself informative:
each of the three mediates a user-visible ranking; interaction prediction
and widget engagement were demonstrably brittle under the poisoning probes
of Section~\ref{sec:poisoning}, and the bandwidth predictor ignores its
state inputs entirely under the adversarial battery
(Section~\ref{sec:adversarial}). A platform that ships runtime-updatable models without signature checks
creates an interface whose intended use---personalization---is at the file
level indistinguishable from tampering.

\subsection{A crash in Apple's own model}
\label{sec:crash}

Loading the system model
\texttt{hand\_gesture\_}\allowbreak\texttt{dynamic\_two\_}\allowbreak\texttt{hand\_5fps.mlmodelc} (shipped inside
the MediaAnalysis hand-gesture framework) through the public
\texttt{MLModel(contentsOf:)} API crashes the process with a segmentation
fault inside \texttt{libBNNS.dylib}, on the CoreML lazy-initialization
queue, during Espresso E5RT ahead-of-time lowering of the MIL program.

\paragraph{Deterministic on production hardware.}
The crash reproduced 4/4 times on a stock iPad Pro (M2, physical Neural
Engine) running iOS~26.6.1, under CPU-only, default, and CPU-and-NE compute
units, and independently on the iOS~26.1 test device as both
\texttt{root} and \texttt{mobile}, where the stack resolves to the BNNS
graph-compilation frame during MIL$\rightarrow$BNNS lowering.
It also reproduces on \textbf{macOS~27.0 (26A428)}, the latest release as of
this writing, through a minimal Swift command-line client loading the same
bundle with \texttt{MLComputeUnits.cpuOnly}---the defect is current, not
historical.

\paragraph{Root cause: a 378-row weight walk with no observed bound check.}
The faulting instruction is a 2-byte fp16 load
(\texttt{ldr h0, [x21, x9]}) in the weight-walking code; the fault address
is in an unmapped gap.
Two twin weight walks (matmul weight tensors of shape
\texttt{fp16[1,12,378,126]} at file offsets 0x89D40 and 0x1D7240) each
iterate \emph{exactly} 378 rows---the row dimension---without checking the
mapped weight-file extent.
Twelve surgical mutations of the model file (M1--M5, M7--M13, plus the
unmodified M0 baseline: truncated
blobs, appended padding, slot-consistent offset moves) bracket each walk to
precisely 378 iterations and show it faults on its first out-of-file row;
files padded past ${\approx}414$~MB, enough for both walks, load without
crashing.
We cannot exclude a checked-but-wrong bound or an overflowed length, but no
effective bound check was observed on either walk.
The walker's \emph{position} is attacker-controllable through the model
data (the blob offsets in the MIL text), while the \emph{absolute} fault
target is ASLR-randomized, the accessor is read-only, and the fault produces
no channel other than the crash---a robustness defect, not memory
corruption or privilege escalation.

\paragraph{Impact.}
Directly, any app that bundles this model crashes itself by loading it---a
self-inflicted fault. The system path is sharper in principle---
\texttt{mediaanalysisd} loads this model on demand for hand-gesture
classification---but we did not observe the daemon performing that load: an
end-to-end test with a hand-containing video did not reproduce a daemon
crash, so daemon-level impact remains a hypothesis. Both demonstrations ran
unsandboxed, so reachability from a third-party app sandbox
(Section~\ref{sec:threat}'s A1) is inferred from the public API surface, not
demonstrated. This is ORCHARD-2026-001; Apple reproduced it and ruled it a
self-crash outside the security boundary (Section~\ref{sec:disclosure}).

\subsection{On-device latency}
\label{sec:latency}

\begin{figure*}[t]
\centering
\includegraphics[width=\textwidth]{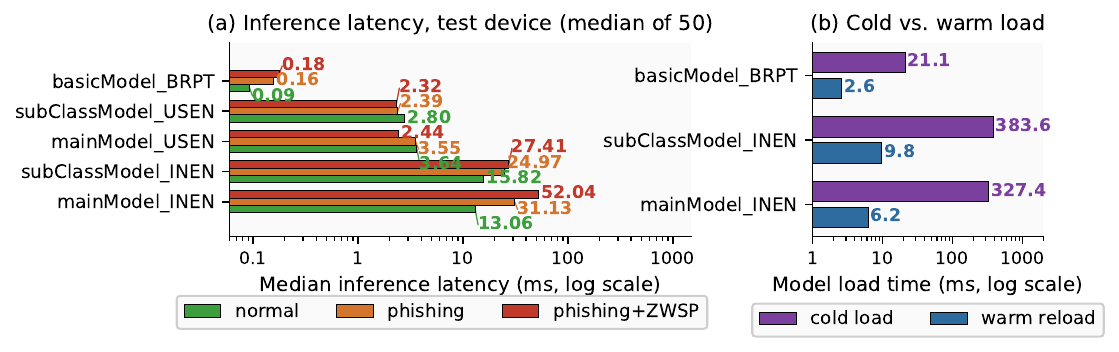}
\caption{Measured on the virtualized iOS~26.1 test device (median of 50
predictions per condition, \texttt{MLComputeUnitsAll}). (a)~Inference latency:
Create~ML text classifiers are dominated by their NLP core---per-stage
instrumentation puts 92.4\% of \texttt{mainModel\_}\allowbreak\texttt{INEN}'s
time in the CPU-side NLP core (tokenizer, dynamic embedding, and
network), not ANE
execution; the tokenized USEN networks and the legacy-format model stay
in the sub-millisecond-to-few-millisecond range. (b)~Cold versus first
warm reload (one later \texttt{subClassModel\_}\allowbreak\texttt{INEN} reload
measured 27.3~ms).}
\label{fig:latency}
\end{figure*}

Earlier drafts reported workstation latencies through
\coremltools{} on macOS; those do not describe the device, whose dominant
cost is the NLP tokenizer and embedding core that desktop runs do not
execute (Figure~\ref{fig:latency}).

\texttt{mainModel\_}\allowbreak\texttt{INEN} takes 13.1~ms per message on ordinary text,
31.1~ms on a phishing sample, and 52.0~ms when the same sample carries
injected zero-width spaces; \texttt{subClassModel\_}\allowbreak\texttt{INEN} takes
15.8--27.4~ms, and the legacy-format \texttt{basicModel\_}\allowbreak\texttt{BRPT} takes
0.09--0.18~ms.
Tokenized USEN networks: 2.3--3.6~ms; cold loads of the three loadable
SMS models: 327.4\text{/}383.6\text{/}21.1~ms, dropping to 2.6--27.3~ms on
subsequent loads (most reloads 2.6--9.8~ms; one
\texttt{subClassModel\_}\allowbreak\texttt{INEN} reload measured 27.3~ms).
Per-stage instrumentation shows where the cost goes: the MLModel wrapper
contributes about 2.0~ms of exclusive time and the inference engine about
2.4~ms, while the NLP core---tokenizer, dynamic embedding, and network---is
53.5~ms inclusive, 92.4\% of the total.
These numbers come from a hook-instrumented run of the live prediction loop
(one \texttt{mainModel\_}\allowbreak\texttt{INEN} classification every 400~ms, 813
predictions over ${\sim}6$~min; 57.9~ms mean instrumented total); the
hooks inflate absolute times, so the shares are the result.

Injecting zero-width characters costs the attacker nothing in
latency terms and costs the \emph{defender} time: the same message
classifies 1.7$\times$ slower with the characters injected (52.0~ms
vs.~31.1~ms)---the characters' value to an attacker is the latency cost and
the benchmark-wide flip rate (65\%, Section~\ref{sec:adversarial}), not a
flip on every message.

\section{Discussion}
\label{sec:discussion}

\subsection{Answers to the research questions}

\textbf{RQ1:} the OS ships more learned behavior than documented, in more
formats and versions, much of it with plaintext parameters.
\textbf{RQ2:} nearly every attacked family has a double-digit worst-case
flip rate, up to 100\% on degenerate models, with two low-flip outliers
among adequately probed families (0.0\% on \texttt{sceneLux}\allowbreak\texttt{B2DItpMLModel}
over 1{,}560 probes; 1.0\% on \texttt{subClassModel\_USEN}); five further
families have too few probes (1--4) to inform this.
\textbf{RQ3:} reading model bytes is trivial; writing them costs one
16-byte edit, with no runtime verification once the seal is defeated.
\textbf{RQ4:} a worst-case SMS classification costs tens of milliseconds,
mostly in the CPU-side NLP core, so load-time verification would not be the
bottleneck.

\subsection{What the adversary classes can actually do}

\emph{A3 (remote)} gets a portable evasion path: zero-width insertion alone
flips 65\% of benchmark messages, the classifiers label 80--90\% of a
ten-message phishing probe as transactional even before manipulation, and
the two-tier architecture provides no second chance; crafted inputs also
amplify latency (Section~\ref{sec:latency}) in a component we did not audit
(the records place the filter in the Messages process).

\emph{A1 (sandboxed app)} gets two no-root primitives, demonstrated
unsandboxed (sandbox reachability inferred and untested): a crash in a
system library through the public model-loading API
(Section~\ref{sec:crash})---a self-crash, though the same compile path is
what the media-analysis daemon drives---and full read
access to any obtainable model's semantics, since weights are plaintext
and compiled bundles recompile with public tooling.

\emph{A2 (local root)} gets the write path only with a jailbroken boot
chain (A2b): one mount suspends the seal and a 16-byte edit flips 7 of 8
probe inputs, with no runtime verification to defeat; a stock-device
runtime root (A2a) cannot rewrite sealed models, and no consumer of daemon
model outputs verifies that the model it calls is the one that shipped.

\subsection{Defense-in-depth gaps}

Three gaps recur.
First, \textbf{no runtime integrity for model data}: \texttt{/System} files
are authenticated in aggregate by the sealed volume, individually by
nothing we could find; once the seal is suspended, arbitrary valid-format
weight edits execute.
A per-file load-time hash would have caught every poisoning payload of
Section~\ref{sec:poisoning} cheaply, and model signing would protect the
updatable path.
Second, \textbf{no input normalization at the model boundary}: the SMS
pipeline accepts zero-width and confusable characters no human would
distinguish; stripping plus compatibility folding addresses the zero-width
variants we measured (hypothesis, no ablation), and homoglyphs need a
confusable-class detector (UTS~\#39~\cite{unicode_tr39}).
Third, \textbf{no validation of model files at compile time}: the walk of
Section~\ref{sec:crash} runs past a data-file boundary with no observed
bound check, on an input that ships with the OS; validating blob extents
against the mapped file removes the fault class.

\subsection{Responsible disclosure}
\label{sec:disclosure}

We filed three reports through the Apple Security Bounty program: the
deterministic crash (ORCHARD-2026-001; tracking ID OE1104416693631), the
internal build-token leakage, and the internal storage-path disclosure
(OE1104416117373). Apple reproduced the crash and replied that \emph{"{}it
only crashes your own application process\ldots it does not cross a security
boundary"}, and declined all three. We agree: the crash is a robustness
defect in the compilation path, not an exploitable vulnerability, and the two
metadata findings are disclosure issues rather than access vectors (all
endpoints tested, all 404; Section~\ref{sec:weights}); we report them as such.
Two further findings we deliberately did not file: the absent runtime weight
integrity, because exploiting it requires local root, and the SMS phishing
baseline, because it is classifier efficacy rather than a boundary crossing.
We publish metadata and analyses; weights are not redistributed
(Appendix~\ref{app:ethics}).

\subsection{Limitations}

\textbf{Scope.} The census covers the harvestable OS versions (iOS~12.5.8,
16.7, 26.0, 26.1; iPadOS~26.0.1; tvOS~26.6; macOS~26.5; no watchOS or
visionOS image was examined), and the 368-plaintext figure comes from a
64~KiB scan per artifact---a lower bound, since some of the 262 quantized
artifacts could embed plaintext outside the sampled regions.

\textbf{Measurement platform.} On-device numbers come from the virtualized
test device (8~GiB, jailbroken rootless); hardware magnitudes may differ,
though seal defeat and absent runtime verification are platform properties
(specific flips are model-specific).

\textbf{Conversion fidelity.} The 155 conversions unlock load and predict,
not full fidelity: they run through the desktop runtime, so agreement with
the device is assumed, and their adversarial results are relative.

\textbf{Benchmark size and proxies.} The SMS benchmark is 20 messages and
the phishing probe 10 templates: flip percentages are exact on those sets,
not population estimates, and the phishing result is the classifier's own
label, not an inbox-routing outcome, which depends on daemon policy we did
not audit. All USEN adversarial numbers use the surrogate character-level
embedding reconstruction (the production tokenizer is not shipped;
Section~\ref{sec:adversarial}), so they characterize the shipped network,
and deployed rates may differ.
Accuracy in Figure~\ref{fig:accuracy} was computed with \coremltools{} on
macOS.

\textbf{Attack prerequisites.} Seal defeat (A2b) requires a jailbroken boot
chain; a stock-device runtime-root attacker is confined to A2a.
Exploitability of the crash beyond denial of service was assessed
statically---read-only accessor, ASLR-random target, no observability---and
the daemon-level impact remains a hypothesis (Section~\ref{sec:crash}):
the direct load still crashes on iOS~27.0, yet a 13.7-hour
\texttt{mediaanalysisd} run on the same updated device executed continuous
\texttt{Espresso} inference without crashing---so the daemon's analysis
cascade never reaches this model's load path.

\section{Related work}
\label{sec:related}

Sun et~al.~\cite{mobile_audit} found app-embedded Android models routinely
extractable; we ask the same of OS-shipped models, whose weights are also
plaintext. Black-box~\cite{papernot_blackbox} and character-level
perturbations~\cite{bad_characters}, systematized by
TextFooler~\cite{textfooler}, established text-classifier fragility to
invisible edits, which we reproduce in a deployed OS feature; confusable
classes are standardized by UTS~\#39~\cite{unicode_tr39} and detected by
\cite{homoglyph_detection}. We recover architecture and weights from the
compiled artifact, where Hua et~al.~\cite{hua_dac} used side channels, and
supply ground truth for the edge pattern Zhou et~al.~\cite{edge_intelligence}
survey.

\section{Conclusion}
\label{sec:conclusion}

We presented, to our knowledge, the first systematic analysis of Apple's
on-device CoreML surface: 1{,}137 artifacts across six platforms, plaintext
weights, no runtime integrity check, and a public-API crash Apple ruled a
self-crash.


\appendix
\section{Ethical Considerations}
\label{app:ethics}

Firmware images were obtained from public distribution channels
(\texttt{ipsw.me} and the Apple Developer downloads); the test device of
Section~\ref{sec:harvesting} is a research virtual device that we own and
operate, jailbroken with a public tool. No user data---messages, photos, contacts, or telemetry---was
accessed, collected, or analyzed at any point; all inputs to models were
synthetic or drawn from public examples.
Poisoning experiments modified system files only on the research device,
were reverted immediately, and were verified byte-identical by SHA-256
against pristine copies, with the volume re-sealed at reboot.
We do not redistribute Apple's model weights, prompts, or tokenizers; we
publish metadata, structural analyses, and reproducibility records.
The deterministic crash (ORCHARD-2026-001), the internal build-token leakage,
and the internal storage-path disclosure were filed with Apple Product Security Security
and declined as findings that do not cross a security boundary. The weight-
integrity and SMS phishing findings were not filed (local root; efficacy). We
report the internal-path disclosure as non-exploitable based on active
testing (Section~\ref{sec:disclosure}) rather than assumption.

\section{Open Science}
\label{app:open}

A reproducibility package accompanies this submission at
\url{https://github.com/jayluxferro/on-device-coreml-security} (anonymously hosted, no tracking).  The package contains:
(i)~\emph{harvest and conversion tooling}---the model-enumeration scripts,
the IPSW mount-based extractor, and the three-stage
\mlmodelc{}$\rightarrow$\mlpackage{} pipeline (loader, MIL rewriter, and
subprocess-isolated verifier);
(ii)~\emph{analysis tooling}---the static analyzer that produced the 1{,}137
records, including the plaintext/quantized/provenance predicates, and the
adversarial harness with all 14 text-evasion generators (32 distinct text
attack types) and the feature-space battery;
(iii)~\emph{evaluation data}---the complete adversarial dataset: 82{,}094
JSONL records in two schemas (82{,}092 campaign rows plus two supplementary
\texttt{resnet-16} style-0 rows). The campaign rows are 82{,}085
\texttt{predict()} calls---77{,}396 carrying an attacked output, 4{,}680
latency-only \texttt{crema}/TFModel calls, five sanity repeats, and four
input-validation rejects---plus seven \texttt{\_\_population\_\_} summary
rows. The package also holds the static-analysis records (so the
plaintext and 262 quantized counts of Table~\ref{tab:static} recompute
exactly), the latency logs, and the poisoning transcripts;
(iv)~\emph{measurements}---the 20-message SMS benchmark and the phishing
template set used in Section~\ref{sec:adversarial} (representative
with full texts living in the released harness and dataset variants); and
(v)~\emph{a crash reproduction}---a minimal SwiftUI CoreML client (its
103-line \texttt{App.swift}) that triggers the \texttt{hand\_gesture}
crash on affected systems through the public model-loading API, given a
user-supplied copy of the model. The directory is packaged for release
from tracked sources only, with an explicit exclusion list
(\texttt{RELEASE\_EXCLUDES.txt}) that strips the bundled Apple model
payload (weight files, MIL text, compiled \texttt{coremldata.bin}) and
Xcode user state (\texttt{xcuserdata}), consistent with the
no-redistribution policy of Appendix~\ref{app:ethics}.
Model files themselves, their weights, and Apple's tokenizer vocabularies
and prompt assets are excluded, since they are Apple's intellectual
property; all dataset records reference models by name and path so
that a researcher with their own device or firmware image can rebuild the
corpus exactly.

\end{document}